\documentclass[aps,prb,twocolumn,showpacs,superscriptaddress]{revtex4-1}

\usepackage{graphicx} % Include figure files

\begin{document}

\title
{Two-dimensional C/BN core/shell structures}

\author{S. Cahangirov}
\affiliation{UNAM-Institute of Materials Science and Nanotechnology, Bilkent University, Ankara 06800, Turkey}
\author{S. Ciraci}
\email{ciraci@fen.bilkent.edu.tr}
\affiliation{UNAM-Institute of Materials Science and Nanotechnology, Bilkent University, Ankara 06800, Turkey}
\affiliation{Department of Physics, Bilkent University, Ankara 06800, Turkey}

\date{\today}

\begin{abstract}
Single layer core/shell structures consisting of graphene as core and hexagonal boron nitride as shell are studied using first-principles plane wave method  within density functional theory. Electronic energy level structure is analysed as a function of the size of both core and shell. It is found that the confinement of electrons in two dimensional graphene quantum dot is reduced by the presence of boron nitride shell. The energy gap is determined by the graphene states. Comparison of round, hexagonal, rectangular and triangular core/shell structures reveals that their electronic and magnetic states are strongly affected by their geometrical shapes. The energy level structure, energy gap and magnetic states can be modified by external charging. The core part acts as a two-dimensional quantum dot for both electrons and holes. The capacity of extra electron intake of these quantum dots is shown to be limited by the Coulomb blockade in two-dimension.
\end{abstract}

 \maketitle

\section{Introduction}
Synthesis of the graphene structure\cite{graphene-synthesis} provided a new perspective to scientific community. This monoatomic layer of carbon atoms is the best media for investigating exciting physical phenomena in two-dimensions (2D). Graphene is a semimetal with unique electronic and mechanical properties. The intensive study on graphene structure rouse interest also in other 2D materials.\cite{prl,ansiklopedi} One of these materials is 2D honeycomb structure of boron nitride (BN). While few atomic layers of BN honeycomb structure was synthesized using micromechanical cleavage technique,\cite{bn-synthesis-few} a single layer was obtained by electron irradiation.\cite{bn-synthesis-single} The mechanical properties of graphene and BN are very similar. They have only 3~\% of lattice mismatch and BN is the stiffest honeycomb structure after graphene.\cite{ansiklopedi} The electronic structure, however, is very different. Graphene is a semimetal, while BN is a wide band gap material.\cite{graphene-gap,bn-gap,can}

Due to their structural similarity and distinct electronic properties hexagonal BN (h-BN) and graphite structures are considered to be good candidates for fabricating composite B/N/C materials offering new functionalities.\cite{bnc} Interesting phenomena like electron confinement, itinerant ferromagnetism and half metallicity was predicted theoretically in quasi-one-dimensional BN/C nanowire, nanotube/nanoribbon superlattices and hybrid BN/C nanoribbons/nanotubes.\cite{bn-c-tub-super,1d-super-lat-rib,hyb-tube-ab-1,gr-rib-emd-bnnr,hyb-tube-ab-2,pruneda,hyb-tube-ab-3} Also nanotubes of BN/C in various patterns were synthesized experimentally.\cite{bn-c-pat-tube-exp,xray-bnc-tube-exp,bn-c-seq-tube-exp} Owing to synthesis of single layer composite structures consisting of adjacent BN and graphene domains,\cite{natmat-bn-c} C/BN core/shell (CS) structures have been a focus of interest. These structures can find novel technological applications in electronics and photonics in the view of the recent realization of nanoscale meshes or periodically patterned nanostructures.\cite{natmat-rubio,meshes,meshru1,meshru2}  Despite the ongoing intensive work on quasi-one-dimensional BN/C structures, 0D and 2D counterparts of these materials are not enough explored yet.

This paper investigates the CS structures of C and BN formed by their commensurate 2D hexagonal lattice. We report the variation of energy gap with respect to geometric parameters of the structures and link these results to quantum confinement phenomena. The shape of the CS structures, in particular whether they have round or rectangular shape, causes crucial effects in the electronic structure. We reveal the nature of the band edge states by analyzing the charge density profiles projected on those states. C/BN CS structure considered here constitutes a planar quantum dot. The effect of charging on these 2D quantum dots is investigated and the role of Coulomb blockade is discussed in detail.

\section{Methods}
Our predictions are obtained from the state-of-the-art first-principles plane wave calculations carried out within the spin-polarized density functional theory (DFT) using the projector augmented wave (PAW) potentials.\cite{paw} The exchange correlation potential is approximated by generalized gradient approximation (GGA) using Perdew-Burke-Ernzerhof functional.\cite{pbe} A plane-wave basis set with kinetic energy cutoff of 500 eV is used. All structures are treated within the supercell geometry; the lattice constants, as well as the positions of all atoms in the supercell are optimized by using the conjugate gradient method, where the total energy and atomic forces are minimized. \cite{vasp} The convergence criterion for energy is chosen as 10$^{-5}$ eV between two steps. In the course of ionic relaxation, the maximum Hellmann-Feynman forces acting on each atom is kept less than 0.01 eV/\AA{}. The vacuum separation between the structures in the adjacent unit cells is taken to be at least 10~\AA. For charged structures this value is at least 15~\AA. Also, the necessary corrections\cite{vasp-chg-1,vasp-chg-2} needed in calculations with charged cells were included in the fashion they are implemented in the simulation package VASP, that we use. To calculate the energy gaps of large graphene patches we use the nearest neighbor tight-binding technique. Here the self energies of carbon atoms are set to zero and the nearest neighbor hopping integral is taken to be $t=2.7$~eV. The hopping integral between hydrogen and carbon atoms at the edges are taken to be zero. Although we mention the quantitative energy gap of the structures throughout the paper, there can be errors due to the lack of the self-energy corrections in the method that we use. We think that, inclusion of these corrections would not result in a qualitative change in the observed trends.

\section{Hexagonal C/BN CS structures: Effects of size}
\begin{figure}
\includegraphics[width=8.4cm]{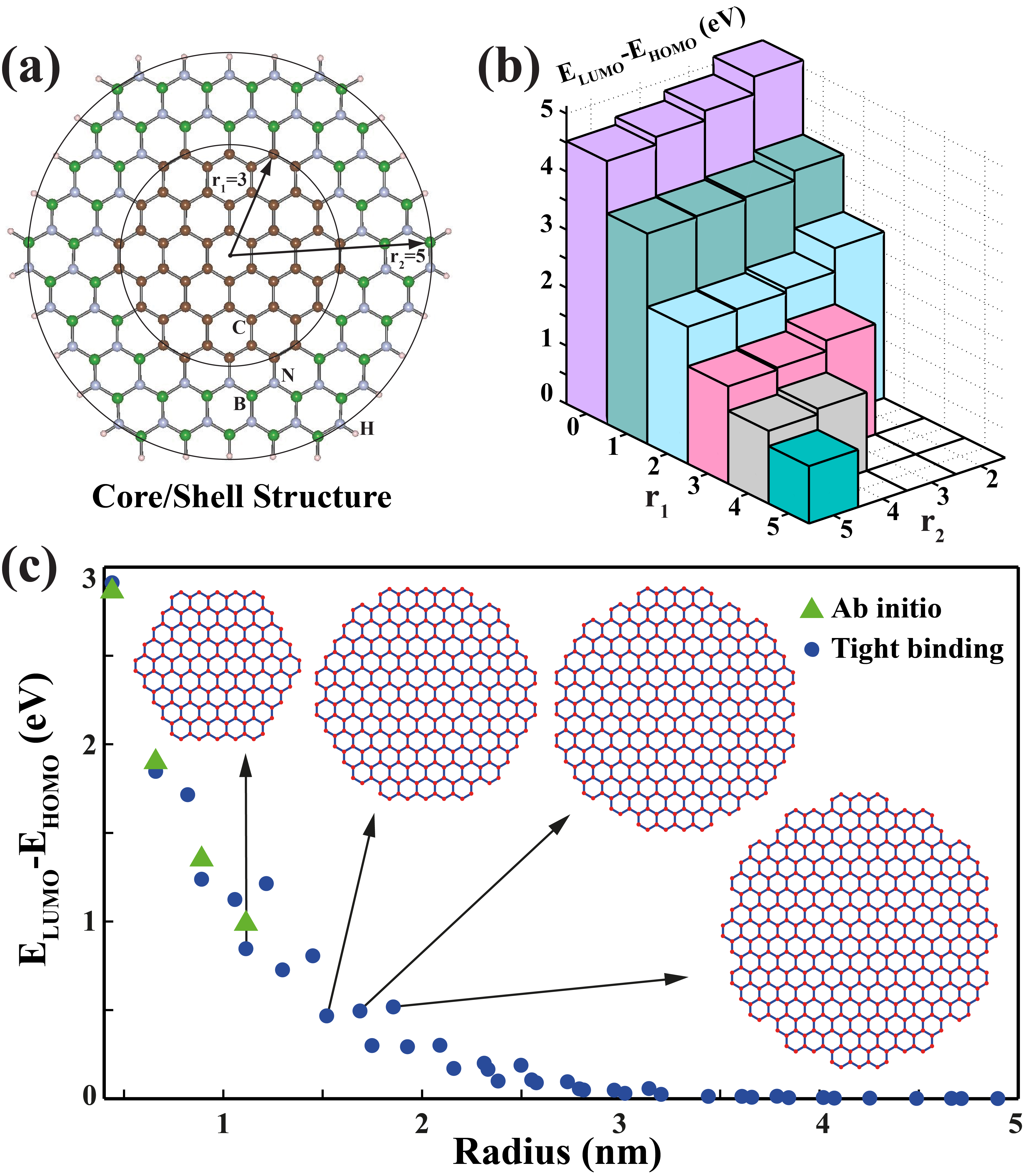}
\caption{(Color Online) (a) Ball and stick representation of single layer C/BN CS structure. The presented structure is denoted as CS($r_1=3$, $r_2=5$). Here radii $r_1$ and $r_2$ are defined as number of hexagonal atomic rings formed by C atoms and total number of hexagonal rings, respectively. Note that, for $r_2 \leq 5$ the atoms enclosed by a circle of this radius make a hexagonal structure. (b) The energy gap between the HOMO and the LUMO levels for hexagonal structures having various $r_1$ and $r_2$ values. (c) Energy gap trends for round graphene patches with various radii and edge patterns calculated from the first-principles and using the nearest neighbor tight-binding approximation. Ball and stick representation of these round structures are presented in the inset.} \label{fig1}
\end{figure}

In the Fig.~\ref{fig1}(a) we present the geometric parameters used to define the hexagonal CS structures. Starting with a single hexagonal atomic structure we call its radius $r=1$ then adding a new hexagonal ring to this structure we get a structure with a radius defined as $r=2$. According to this definition we have investigated all hexagonal CS structures having total radius from $r_2=2$ to $r_2=5$ and inner radius from $r_1=0$ to $r_1=r_2$. This means that we also investigate structures composed of only C and only BN, which provide data for comparison. Note that, the low coordinated edge atoms are passivated by H atoms in all structures.

Figure~\ref{fig1}(b) presents the calculated energy difference between the highest occupied (HOMO) and the lowest unoccupied (LUMO) molecular orbitals for all hexagonal CS structures investigated here. Due to the threefold symmetry of these structures, both the HOMO and the LUMO states are doubly degenerate. One can find several trends in Fig.~\ref{fig1}(b). For instance, $r_1=0$ corresponds to hexagonal BN patches. As the radius $r_2$ of these structures is varied from 2 to 5 the energy gap is reduced by 0.6 eV. For comparison we can look at structures with $r_1=r_2=2$ to 5 which corresponds to hexagonal C patches. This time the energy gap is lowered by 2.9 eV. This large difference as compared to BN occurs because in the graphene structure the $\pi$-orbitals of adjacent carbon atoms have the same energy and thus interact strongly. This leads to dispersive bands. In BN structure, however, $\pi$-orbitals of adjacent B and N atoms have large energy difference and have weaker interaction. This results in rather flat bands. In fact, the charge density projected on the HOMO states of BN structure is composed of weakly interacting (second nearest neighbor) $\pi$-states of N atoms. For graphene, the band edge states are composed of strongly interacting $\pi$-states of neighboring C atoms. Since the energy gap of BN and C structures with a certain finite size corresponds to certain folding of periodic band profiles, the dispersion of these bands are mirrored in the LUMO-HOMO gap variation of the finite structures with their size.

Now let $r_1=2$, when $r_2$ goes from 2 to 3 we add a BN layer to hexagonal C structure. $r_2=4$ and 5 correspond to adding further BN layers to this CS structure. By adding one BN layer the energy gap decreases by 0.4 eV, however adding two more layers the energy gap decreases further by only 0.1 eV. Either surrounded by vacuum or BN layers, the graphene structure acts as a quantum well. Thus, the band edge states are confined in the graphene region. When the graphene structure is surrounded by the vacuum, these states are confined in higher potential barrier than when it was surrounded by BN structure. Buffering BN layer reduces the confinement strength, thereby reducing the energy gap. However this reduction in energy gap is low, compared to that when $r_1$ parameter of CS structure is increased. This again shows that the band edge states are determined by the core part, and the shell part has a minor effect on them.

We have to note that, the structures mentioned so far are very small and hard to be synthesized experimentally, but larger structures compatible with experiment cannot be treated from the first-principles. For example, the size of BN domains in the recently synthesized BN/graphene heterostructure is estimated to range from 2~nm to 42~nm.\cite{natmat-bn-c} Also, graphene nanoflakes with sizes in the order of 10~nm have been synthesized by both bottom-up\cite{rrrl} and top-down\cite{rrra} approaches.

Here we extrapolate our results obtained for the smaller structures using a single parameter nearest neighbor tight-binding calculation. Since the band edge states are determined by the core graphene structure, we have only performed tight-binding calculations of graphene patches saturated by hydrogen atoms at the edges. The results of these calculations are presented in Fig.~\ref{fig1}(c). One can see that, for smaller structures the tight binding and ab initio calculations are in a good agreement. One can clearly see that the energy gaps decrease as the radii of the structures increase, which is due to the quantum size effect. However, this trend is sometimes breaks down due to the dominating effect of edge termination in smaller patches. This kind of behavior is highlighted in Fig.~\ref{fig1}(c). Here one can see the increase in the energy gap with increasing radius, which is accompanied with different edge patterns shown in the inset. As the radius increases further, the effect of the edges ceases and the energy gap approaches zero. In fact, the calculated energy gap for round graphene patches having radii more than 4.5~nm is lower than 1~meV.

Figure~\ref{fig2} presents the energy level diagram for core, shell and CS structures individually. This figure does not include a band offset information; the first two columns are presented only for the sake of comparison. As mentioned in the previous paragraph, adding a shell reduces confinement of electrons in the core structure. This squeezes the energy levels slightly, but still the HOMO and the LUMO states of core structures are present in the CS structure, because the shell structure by itself has a large energy gap.  Charge densities of the HOMO and the LUMO shown in the Fig.~\ref{fig2} clearly corroborate this fact. One can see the bonding and antibonding states of the doubly degenerate HOMO and LUMO confined in the core part. The nondegenerate state below the HOMO is confined at the interface of the core and the shell structure. However, this state is also attributed to the core structure, because core structures by themselves possess this state as an edge state. Moving deeper to the filled molecular orbitals one can also find states spread both on the core and shell parts. Also as shown in the bottom charge density plot of Fig.~\ref{fig2}, deep down the filled orbitals there are states which are confined in the shell part.

\section{Round, rectangular and triangular C/BN CS structures: Effects of geometry}
 We now investigate the effect of shape on the electronic properties of the CS structures. Here we compare the energy levels and charge densities of the band edge states of round, rectangular and triangular CS structures with that of the hexagonal structures mentioned in the previous section. We start with the analysis of the large (C$_{204}$B$_{123}$N$_{123}$H$_{54}$) structure, which is composed of 504 atoms and mimics a round CS structure. This way we can contrast hexagonal and round CS structures. The core radius of this structure is 12.7~\AA{} while the total radius is 19.4~\AA. Unlike the hexagonal structures having linear boundaries composed of zigzag edges, this structure has round boundaries with armchair and zigzag edges at both core and shell parts. Similar to hexagonal structures, the round structure also has threefold symmetry which results in the doubly degenerate band edge states. These states are confined in the core part of the round structure. The energy gap between its HOMO and LUMO states is calculated to be 0.77 eV. Moreover, as hexagonal structures, the round CS structures also have a nondegenerate state below the HOMO level, which is confined at the interface of the core and shell parts. Note that, both hexagonal and round structures have equal number of C atoms in each sublattice and, as expected by Lieb's theorem,\cite{lieb} have a nonmagnetic ground state.

\begin{figure}
\includegraphics[width=8.4cm]{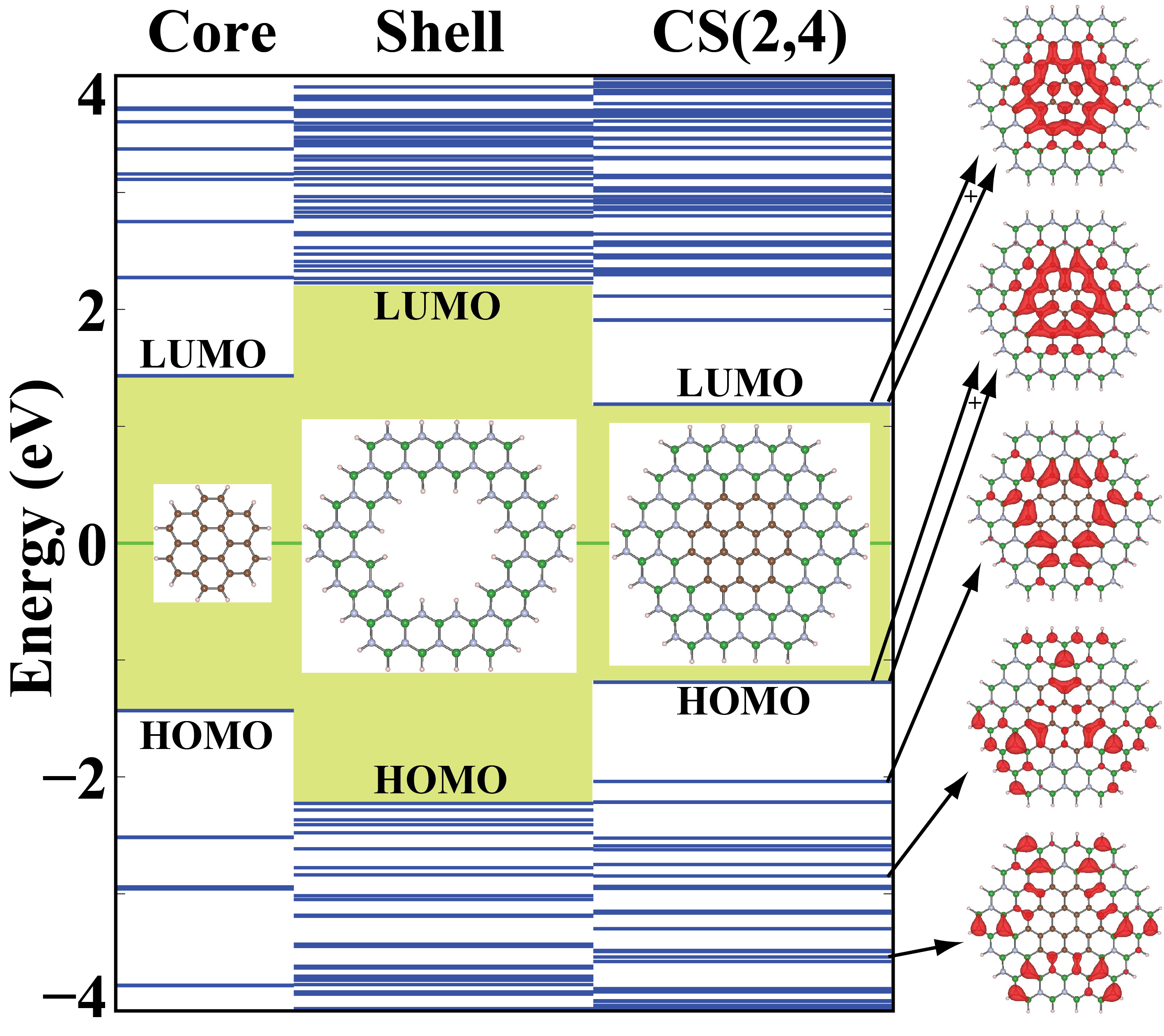}
\caption{(Color Online) Energy level diagram for fully relaxed hexagonal structure of C atoms, hexagonal shell structure of BN and CS structure which is formed by the combination of the first two structures. Isosurface charge densities projected on specific states of CS(2,4) structure. Single arrows correspond to the nondegenerate states. Charge densities of the doubly degenerate band edge states are shown by two arrows. The `+' sign between these two arrows denotes that the plotted charge density profile is the average values of two degenerate states. All plots have the same isosurface values. Zero of energy is set to the middle of the gap between HOMO and LUMO. The energy gap region is shaded.}
\label{fig2}
\end{figure}

\begin{figure}
\includegraphics[width=8.4cm]{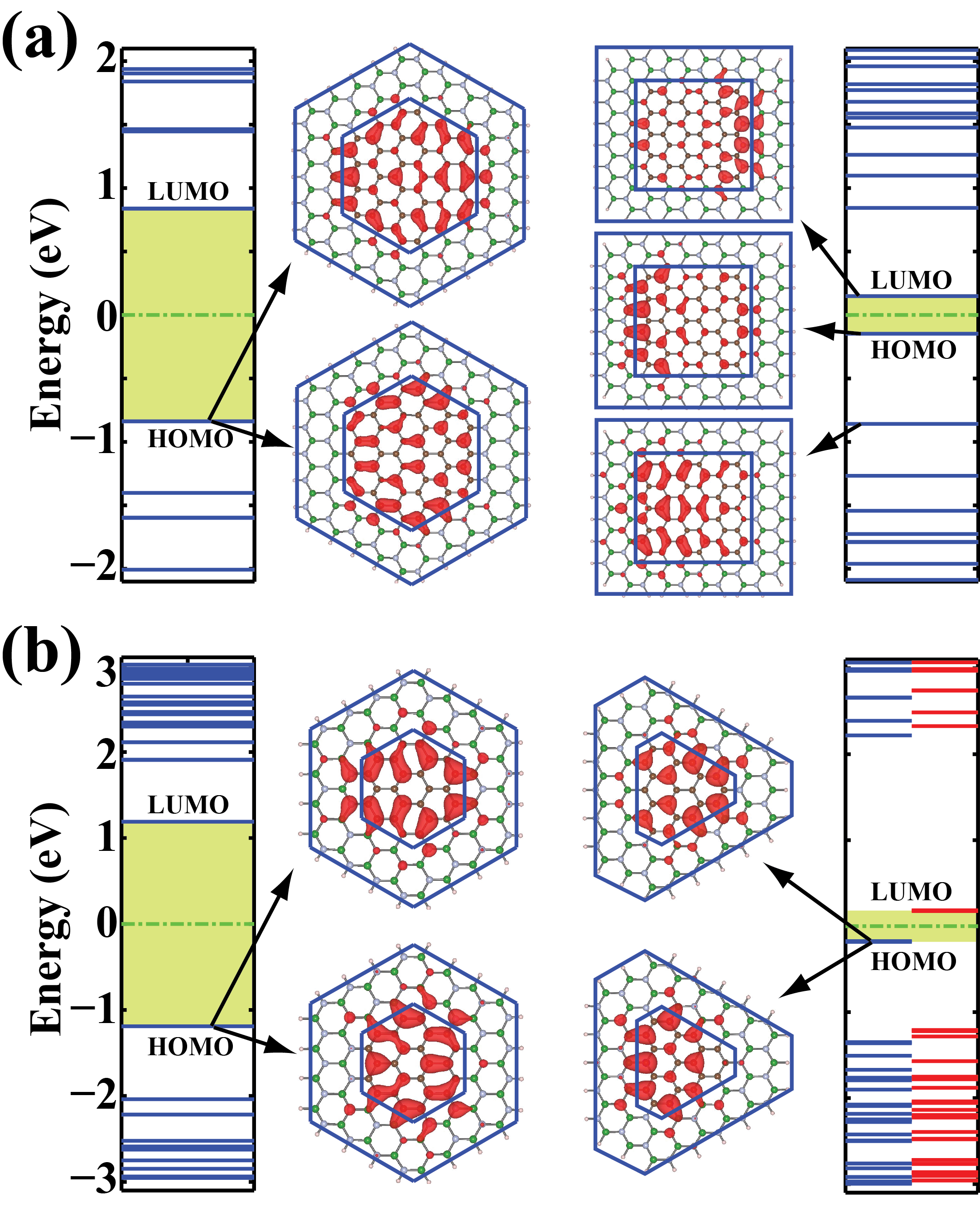}
\caption{(Color Online) (a) Comparison of hexagonal (C$_{54}$B$_{48}$N$_{48}$H$_{30}$) and rectangular (C$_{54}$B$_{46}$N$_{46}$H$_{32}$) CS structures. C and BN regions are delineated on the isosurface charge density plots. Band edge states of hexagonal structure is doubly degenerate, while in rectangular structure this degeneracy is lifted. (b) Comparison of hexagonal (C$_{24}$B$_{36}$N$_{36}$H$_{24}$) and triangular (C$_{22}$B$_{36}$N$_{36}$H$_{24}$) CS structures. Split majority and minority spin states of triangular structure are shown side by side. Band edge states of both hexagonal and triangular structures are doubly degenerate. Zero of energy is set at the middle of the energy gap between HOMO and LUMO and is shown by a dashed dotted lines. The energy gap region is shaded.} \label{fig3}
\end{figure}

Figure~\ref{fig3} presents a comparison of the electronic energy levels of hexagonal, rectangular and triangular CS structures. To explore the effect of the shape we try to compare structures having similar stoichiometry. Energy level diagram and charge densities projected to HOMO states of hexagonal (C$_{54}$B$_{48}$N$_{48}$H$_{30}$) and rectangular (C$_{54}$B$_{46}$N$_{46}$H$_{32}$) CS structures are presented in Fig.~\ref{fig3}(a). Here we again see that the hexagonal CS structure have twofold degenerate HOMO states confined in the core part. HOMO states of rectangular CS structure are also confined in the core region, but upon the removal of the threefold symmetry the twofold degeneracy is lifted. One of the band edge states of rectangular structure remains at the same level as that of the hexagonal structure. This is attributed to nearly the same area of core parts which enclose the same number of carbon atoms and also that the charge density of this state is spread over the whole rectangular core part. The other band edge state, however, is shifted close to the Fermi level, thereby decreasing the band gap. The charge density of this state is confined at the zigzag edges of the core part of the rectangular structure. We have also calculated the electronic structure of the (C$_{54}$H$_{20}$) structure which is composed of only core part of presented rectangular CS structure, namely rectangular core structure. Compared to rectangular CS structure, the states that are spread over whole rectangular core structure are located further apart from the Fermi level. In the previous section, similar effect was mentioned about the HOMO-LUMO gap of the hexagonal core structure, compared to hexagonal CS structure with the same core size. Conversely, HOMO and LUMO states of rectangular core structure are closer (0.03~eV) to Fermi level than those of the rectangular CS structure (0.15~eV). In fact, the HOMO and the LUMO states of the rectangular core structure are very similar and confined at both zigzag edges. Adding the shell, which consists of the dipolar BN, breaks the symmetry of the effective electronic potential of the rectangular core structure. The resulting effective potential of the rectangular CS structure is deeper in the left hand side, which further confines the bonding HOMO states towards left and the antibonding LUMO states towards right. This increase in the confinement strength explains the increase in the energy gap of the rectangular CS structure compared to that of the rectangular core structure. Rectangular CS structures presented here have the same number of C atoms in each sublattice and thus have total magnetic moment of zero.

In Fig.~\ref{fig3}(b) the energy level diagram and charge densities of HOMO of hexagonal (C$_{24}$B$_{36}$N$_{36}$H$_{24}$) and triangular (C$_{24}$B$_{36}$N$_{36}$H$_{24}$) CS structures are presented. Note that, the hexagonal structure presented in this figure is also CS(2,4). Thus, the sum of the charge densities of the twofold degenerate band edge states of this structure can be compared to that in Fig.~\ref{fig2}. Since the threefold symmetry is preserved in triangular CS structures, the band edge states are again doubly degenerate. But this time the difference between the number of atoms in each sublattice of the core structure is nonzero, being equal to two. Unpaired $\pi$-orbitals result in the breaking of the spin degeneracy and hence the structure gains a net magnetic moment of 2~$\mu_B$. The difference in the number of atoms in each sublattice is proportional to the perimeter of the triangular core structures having zigzag edges. Thus, the magnitude of the magnetization is expected to increase with increasing size of the triangular core structures.\cite{lieb,rrrg,lieb-hasan} Because of the comparable core area, the HOMO-LUMO gap of the majority spins of the triangular CS structure is close to the gap of the hexagonal CS structure. However, the overall HOMO-LUMO gap of the triangular CS structure is diminished to 0.4~eV as compared to 2.4~eV gap of the hexagonal CS structure. Actually, the spin degenerate tight-binding and DFT calculations give zero energy gap for triangular core structures. Spin polarized tight-binding calculation\cite{rrre} of triangular core structure results in 0.8~eV energy gap, and our DFT calculation gives 0.6~eV, while the net magnetic moment is 2~$\mu_B$. Note that, the energy gap, which is caused by the magnetization, is decreased upon the inclusion of the shell structure. Similarly, while the magnetic state of the triangular core structure is favourable over the nonmagnetic state by 0.27~eV, in the triangular CS structure this energy difference is diminished to 0.15~eV. From these trends we deduce that, including the shell structure weakens the magnetization of the system, while preserving the total magnetic moment.

\begin{figure}
\includegraphics[width=8.4cm]{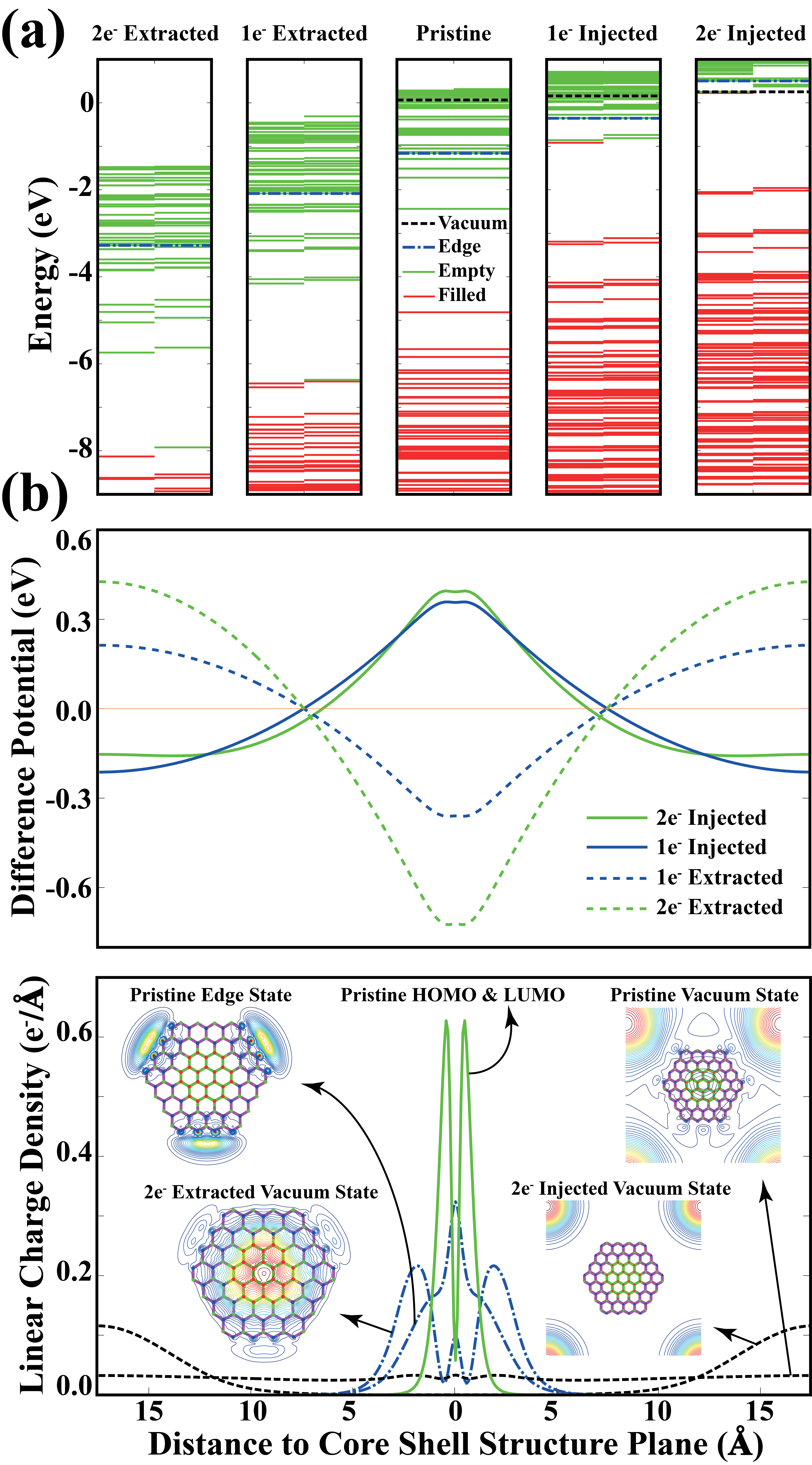}
\caption{(Color Online) (a) Energy level diagram for pristine and charged CS(2,4) structure. The minority and majority spin levels are shown at the left and right side of the same plot, respectively. The filled states are shown by dark (red) and the empty states are shown by light (green) lines. The energy level of edge and vacuum states mentioned in the text are shown by dashed dotted (blue) and dashed (black) lines, respectively. In all plots except the two electron injected case, zero of energy is set to the local potential value at the farthest place from the surface. The reason for this exception and the definition of zero energy for this plot is explained in the text. (b) Top panel: planarly averaged local potential difference between charged and pristine structures, plotted along the axis perpendicular to the surface. Bottom panel: planarly averaged linear charge density for specific states, plotted along the axis perpendicular to the structures. Contour plots of specific states averaged along this axis are shown on the honeycomb structure.}
\label{fig4}
\end{figure}

\section{Effects of charging}
From previous sections one can deduce that the CS structures of C/BN can be used as a quantum dot with core structure behaving as a quantum well for both electrons and holes. One can trap both positive and negative charges in these quantum dots.\cite{rrrm,rrrn} To explore the effect of charging we have calculated the electronic structure of CS(2,4) structure when one or two electrons are added to or removed from the system.

Upon the removal of one electron the CS(2,4)$^{+1}$ structure is formed. Minority spin part of one of the doubly degenerate HOMO is emptied and the system gains a magnetic moment of 1~$\mu_{B}$. The CS(2,4)$^{+2}$, which is formed upon the removal of the second electron has a doubly degenerate HOMO composed of majority spins. This situation gives rise to a total magnetic moment of 2~$\mu_{B}$. The LUMO is composed of doubly degenerate minority spin states located 0.2~eV above the HOMO. When one electron is added to pristine CS(2,4) structure the energy of the system lowers and a magnetic moment of 1~$\mu_{B}$ is acquired. However, upon the injection of the second electron, the energy of the system is not lowered further, but increased. This is interpreted as the electron intake of the CS(2,4) structure being limited to only one. The forthcoming analysis corroborates this conclusion.

Figure~\ref{fig4}(a) presents the energy level diagram for pristine and charged CS(2,4) structure with one or two electrons added or removed. One can see the levels shifting up as the electron is injected in the system. This is mainly attributed to increase in the  Coulomb repulsion term on the electrons as the number of electrons is increased in the system. The magnitude of the shift is $\sim$1.6~eV per injected electron for both HOMO and LUMO levels. This can be explained by similar charge density distribution in HOMO and LUMO, both of which are confined in the core part. This shift sets a limit to the electron intake of the CS structure. As seen in Fig.~\ref{fig4}(a), the LUMO of the CS(2,4)$^{-1}$ structure is 0.9~eV below the vacuum level. This is the first available state for the second electron injection, but due to the shift of approximately 1.6~eV, the state will move above the vacuum level. This means that the second electron will stay in the vacuum, rather than filling the LUMO level of CS(2,4)$^{-1}$. This limitation of electron intake is a reminiscent of Coulomb blockade phenomena in 2D quantum dots. \cite{rrrc}

Zero of energy in the level diagram of CS(2,4)$^{-2}$ is not set to the local potential value at the most distant point to the CS structure in the unit cell. This is because the second electron injected to the system fills the vacuum state, which is a state that is mainly spread far from the CS structure. Since such state is filled, the local potential at the vacuum region becomes ill-defined. This is seen in the top panel of Fig.~\ref{fig4}(b), where we present the planarly averaged plot of local potential of charged structure minus that of the pristine structure. Here one can see that extracting two electrons from the system doubles the change of local potential profile formed upon extraction of only one electron. This linearity holds also in the case of injecting one electron to the system, where the change of local potential profile has the mirror symmetry of that of extracting one electron. However, injecting second electron does not give rise to practically any change in the local potential calculated for single electron injected case. Here the second electron is escaped into the vacuum, far from the CS atomic plane, thereby reducing the depth of potential, which otherwise should be two times deeper than that of one electron injected case. Also the height of local potential difference averaged on the CS plane is not doubled because the second electron is not confined in the core part of CS structure as the first one does. For the reasons stated here, we have not set zero of energy according to the ill-defined local potential of CS(2,4)$^{-2}$ structure. Instead, we have extrapolated the position of slow varying vacuum level state of CS(2,4) and CS(2,4)$^{-1}$ structures to set the vacuum level of CS(2,4)$^{-2}$ structure.

The bottom panel of Fig.~\ref{fig4}(b) presents the planarly averaged linear charge density for specific states.\cite{lchg} The linear charge densities of pristine HOMO and LUMO states have a node at the plane of the CS structure with two symmetric rapidly decaying gaussians. This indicates the $\pi$ nature of these states. The linear charge density of the vacuum state of pristine structure have nearly constant profile which indicates that this state barely interacts with the neutral CS structure. As seen in the charge density contour plots presented by inset, this state is spread throughout the vacuum part of the unit cell. The energy of vacuum states is indicated by dashed black lines in Fig.~\ref{fig4}(a). Below these dashed lines there are no states with charge density spread throughout the vacuum. Note that, the energy of vacuum states are very close to the vacuum level determined by the local potential (zero of energy). When two electrons are injected to the system, the second electron fills the vacuum state. This time the CS structure is negatively charged and thus pushes the vacuum state away. This is clearly seen in both linear charge density and planar contour plot of this state.

Below the vacuum state of pristine structure there is an empty state which is localized around the CS structure but with a significant extension in the vacuum region. Such states were investigated in bulk graphite\cite{pos-gr} and hexagonal BN\cite{pos-bn} structures and were referred as interlayer states. Nevertheless, they were then observed also in isolated graphene\cite{girit-g} and BN\cite{cohen-b} sheets. In fact, this state forms the conduction band edge of bulk and 2D hexagonal structure of BN and makes these structures an indirect gap material.\cite{cohen-b} Carbon\cite{saito-cnt} and BN\cite{cohen-bnt} nanotubes also possess unoccupied free-electron-like states originating from `interlayer' state of their 2D counterparts.\cite{image} In the bottom panel of Fig.~\ref{fig4}(b) we present the linear charge density and planar contour plots of `interlayer' states of pristine and $+2e$ charged CS structures. We refer to these states as edge states. The location of edge states are delineated by dashed dotted lines in Fig.~\ref{fig4}(a). One can see that these states also shift up upon injection of electron to the system. The magnitude of shift is $\sim$1.0~eV per injected electron, which is lower compared to that of the HOMO and LUMO, because this state has large extension in the vacuum region. The vacuum state itself shifts by only $\sim$0.1~eV upon injection of one electron.

We can readily say that the number of electrons that can be injected in quantum dots formed by C/BN CS structures are mainly limited by electron-electron Coulomb interactions. This is, so called, Coulomb blockade in 2D. It should be noted that, the charging capacity of CS structures can be increased by increasing their size. The larger the core part of the CS structure the more states will be available. Also since electrons will be spread in a larger core area, the Coulomb interaction between electrons will decrease.\cite{rrrc} This will decrease the magnitude of the shift in the states due to electron injection. We have performed the electron injection analyses on larger structures to test these trends. Upon injection of one electron into both hexagonal and rectangular CS structures, presented in the Fig.~\ref{fig3}(a), the energy levels were shifted up by $\sim$1.1~eV. In CS(2,4) structure, this shift is $\sim$1.5~eV, which corroborates the trend. Our calculations also show that, the intake capacity of both hexagonal and rectangular CS structures presented in Fig.~\ref{fig3}(a) is two electrons. This is also expected, since the core part of these structures are larger than that of the CS(2,4) structure. We should note that, the shape of the CS structures have minor effect on the magnitude of the shift of the energy levels upon charging and thus have minor effect on the electron intake capacity.

\section{Discussions and Conclusions}
We presented a theoretical analysis of a single layer, composite structure consisting of graphene as core and hexagonal BN as shell, which can make a pseudomorphic heterostructure with a 3\% lattice mismatch. Graphene being a semimetal or semiconductor and BN being a wide band gap insulator the graphene/BN CS structure can function as a quantum dot. Recent studies \cite{natmat-bn-c} reporting the synthesis of composite structures consisting of adjacent hexagonal BN and graphene domains herald the realization of graphene/BN CS structures. In this study, we examined the effects of size, geometry and charging on the electronic and magnetic properties of graphene/BN CS structures. Unexpectedly, the confinement of states in graphene core is relaxed by the presence of BN shell. The band gap decreases with increasing the size of the graphene core. The energy level structure, HOMO-LUMO gap and magnetic state of C/BN CS structures can be tuned by engineering their size and shape, which may lead to interesting applications. Another recent progress in nanotechnology reporting the fabrication of periodically repeating patterns, or simply nanomeshes,\cite{meshes} can be combined with the fabrication of C/BN CS structures to realize arrays of graphene domains in BN layer. Here BN layer can play the role of a dielectric holding multiple and periodic domains of graphene. In visa versa BN dielectric domains in semimetallic graphene layer can be an interesting structure for photonic applications.

\section{Acknowledgement}
This work is partially supported by TUBA. We thank the DEISA Consortium (www.deisa.eu), funded through the EU FP7 project RI-222919, for support within the DEISA Extreme Computing Initiative. Also, part of the computations have been provided by UYBHM at Istanbul Technical University through a Grant No. 2-024-2007. We thank Hasan \c Sahin for fruitful discussions.


\begin{thebibliography}{15}

\bibitem{graphene-synthesis}
K. S. Novoselov, A. K. Geim, S. V. Morozov, D. Jiang, Y. Zhang, S. V. Dubonos, I. V. Grigorieva, and A. A. Firsov, Science \textbf{306}, 666 (2004).

\bibitem{prl}
S. Cahangirov, M. Topsakal, E. Akt\"urk, H. \c Sahin, and S. Ciraci, Phys. Rev. Lett. \textbf{102}, 236804 (2009).

\bibitem{ansiklopedi}
H. \c Sahin, S. Cahangirov, M. Topsakal, E. Bekaroglu, E. Akt\"urk, R. T. Senger, and S. Ciraci, Phys. Rev. B \textbf{80}, 155453 (2009).

\bibitem{bn-synthesis-few}
D. Pacil\'e, J. C. Meyer, \c C. \"O. Girit, and A. Zettl, Appl. Phys. Lett. \textbf{92}, 133107 (2008).

\bibitem{bn-synthesis-single}
C. Jin, F. Lin, K. Suenaga, and S. Iijima, Phys. Rev. Lett. \textbf{102}, 195505 (2009).

\bibitem{graphene-gap}
K. Novoselov, Nature Mater. \textbf{6}, 720 (2007).

\bibitem{bn-gap}
K. Watanabe, T. Taniguchi, and H. Kanda, Nature Mater. \textbf{3}, 404 (2004).

\bibitem{can}
C. Ataca and S. Ciraci, Phys. Rev. B \textbf{82}, 165402 (2010).

\bibitem{bnc}
M. Kawaguchi, Adv. Mater. \textbf{9}, 615 (1997).

\bibitem{bn-c-tub-super}
J. Choi, Y. H. Kim, K. J. Chang, and D. Tom\'anek, Phys. Rev. B \textbf{67}, 125421 (2003).

\bibitem{1d-super-lat-rib}
H. Sevin\c cli, M. Topsakal, and S. Ciraci, Phys. Rev. B \textbf{78}, 245402 (2008).

\bibitem{hyb-tube-ab-1}
A. Du, Y. Chen, Z. Zhu, G. Lu, and S. C. Smith, J. Am. Chem. Soc. \textbf{131}, 1682 (2009).

\bibitem{gr-rib-emd-bnnr}
Y. Ding, Y. Wang, and J. Ni, Appl. Phys. Lett. \textbf{95}, 123105 (2009).

\bibitem{hyb-tube-ab-2}
Z. Y. Zhang, Z. Zhang, and W. Guo, J. Phys. Chem. C \textbf{113}, 13108 (2009).

\bibitem{pruneda}
J. M. Pruneda, Phys. Rev. B \textbf{81}, 161409(R) (2010).

\bibitem{hyb-tube-ab-3}
B. Huang, C. Si, H. Lee, L. Zhao, J. Wu, B. Gu, and W. Duan, Appl. Phys. Lett. \textbf{97}, 043115 (2010).

\bibitem{bn-c-pat-tube-exp}
W. L. Wang, X. D. Bai, K. H. Liu, Z. Xu, D. Golberg, Y. Bando, and E. G. Wang, J. Am. Chem. Soc. \textbf{128}, 6530 (2006).

\bibitem{xray-bnc-tube-exp}
S. Y. Kim, J. H. Park, H. C. Choi, J. P. Han, J. Q. Hou, and H. S. Kang, J. Am. Chem. Soc. \textbf{129}, 1705 (2007).

\bibitem{bn-c-seq-tube-exp}
S. Enouz, O. St\'ephan, J.-L. Colliex, and A. Loiseau, Nano Lett. \textbf{7}, 1856 (2007).

\bibitem{natmat-bn-c}
L. Ci, L. Song, C. Jin, D. Jariwala, D. Wu, Y. Li, A. Srivastava, Z. F. Wang, K. Storr, L. Balicas, F. Liu, and P. M. Ajayan, Nature Mater. \textbf{9}, 430 (2010).

\bibitem{natmat-rubio}
A. Rubio, Nature Mater. \textbf{9}, 379 (2010).

\bibitem{meshes}
J. Bai, X. Zhong, S. Jiang, Y. Huang, and X. Duan, Nature Nanotech. \textbf{5}, 190 (2010).

\bibitem{meshru1}
A. Goriachko and H. Over, Zeitschrift f\"ur Physikalische Chemie \textbf{223}, 157 (2009).

\bibitem{meshru2}
H. Ma, M. Thomann, J. Schmidlin, S. Roth, M. Morscher and T. Greber, Frontiers of Physics in China \textbf{5}, 387 (2010).

\bibitem{paw}
P. E. Bl\"ochl, Phys. Rev. B \textbf{50}, 17953 (1994).

\bibitem{pbe}
J. P. Perdew, K. Burke, and M. Ernzerhof, Phys. Rev. Lett. \textbf{77}, 3865 (1996).

\bibitem{vasp}
G. Kresse and J. Furthmuller, Phys. Rev. B \textbf{54}, 11169 (1996).

\bibitem{vasp-chg-1}
G. Makov and M. C. Payne, Phys. Rev. B \textbf{51}, 4014 (1995).

\bibitem{vasp-chg-2}
J. Neugebauer and M. Scheffler, Phys. Rev. B \textbf{46}, 16067 (1992).

\bibitem{rrrl}
X. Yan, X. Cui, B. Li, and L. Li, Nano Lett. \textbf{10}, 1869 (2010).

\bibitem{rrra}
L. C. Campos, V. R. Manfrinato, J. D. Sanchez-Yamagishi, J. Kong, and P. Jarillo-Herrero, Nano Lett. \textbf{10}, 2600 (2009).

\bibitem{lieb}
E. H. Lieb, Phys. Rev. Lett. \textbf{62}, 1201 (1989).

\bibitem{rrrg}
W. L. Wang, S. Meng, and E. Kaxiras, Nano Lett. \textbf{8}, 241 (2008).

\bibitem{lieb-hasan}
H \c Sahin, R. T. Senger and S. Ciraci J. Appl. Phys. \textbf{108}, 074301 (2010).

\bibitem{rrre}
O. V. Yazyev, Rep. Prog. Phys. \textbf{73} 056501 (2010).

\bibitem{rrrm}
A. D. G\"u\c cl\"u, P. Potasz, O. Voznyy, M. Korkusinski, and P. Hawrylak, Phys. Rev. Lett. \textbf{103}, 246805 (2009).

\bibitem{rrrn}
J. M. Pereira , V. Mlinar, F. M. Peeters, and P. Vasilopoulos, Phys. Rev. B \textbf{74}, 045424 (2006).

\bibitem{rrrc}
M. Ezawa, Phys. Rev. B \textbf{77}, 155411 (2008).

\bibitem{lchg}
The planarly averaged linear charge density is defined as $\lambda (z) = \int \int |\Psi(x,y,z)|^2 dx dy$, where $z$ is perpendicular to the CS structure.

\bibitem{pos-gr}
M. Posternak, A. Baldereschi, A. J. Freeman, and E. Wimmer, Phys. Rev. Lett. \textbf{52}, 863 (1984).

\bibitem{pos-bn}
A. Catellani, M. Posternak, and A. Baldereschi, H. J. F. Jansen and A. J. Freeman, Phys. Rev. B \textbf{32}, 6997 (1985).

\bibitem{girit-g}
D. Pacile, M. Papagno, A. Fraile Rodriguez, M. Grioni, L. Papagno, \c C. \"O. Girit, J. C. Meyer, G. E. Begtrup, and A. Zettl, Phys. Rev. Lett. \textbf{101}, 066806 (2008).

\bibitem{cohen-b}
X. Blase, A. Rubio, S. G. Louie, and M. L. Cohen, Phys. Rev. B \textbf{51}, 6868 (1995).

\bibitem{saito-cnt}
S. Okada, A. Oshiyama, and S. Saito, Phys. Rev. B \textbf{62}, 7634 (2000).

\bibitem{cohen-bnt}
X. Blase, A. Rubio, S. G. Louie, and M. L. Cohen, Europhys. Lett. \textbf{28}, 335 (1994).

\bibitem{image}
V. M. Silkin, J. Zhao, F. Guinea, E. V. Chulkov, P. M. Echenique, and H. Petek, Phys. Rev. B \textbf{80}, 121408(R) (2009).

\end{thebibliography}
\end{document}